\documentclass[a4paper,11pt]{article}
\usepackage{pos}
\DeclareMathOperator{\erfc}{erfc}

\usepackage{orcidlink}

\title{Easier parameter scans with hmcdj }

\author*[a,b,c]{Alexis Verney-Provatas\,\orcidlink{0000-0002-3132-9621}}
\author[a,b]{Ed Bennett\,\orcidlink{0000-0002-1678-6701}}
\author[b]{Gaurav Ray\,\orcidlink{0000-0003-1956-7659}}

\onbehalf{\\for the TELOS Collaboration \vspace{8pt}\\ \hfill\href{https://telos-collaboration.github.io}{ \includegraphics[height=1cm]{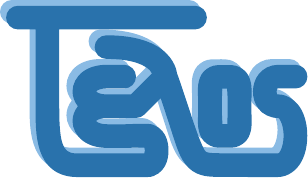} } \hfill{}}

\affiliation[a]{Centre for Quantum Fields and Gravity, Faculty of Science and Engineering, Swansea University,\\
Singleton Park, SA2 8PP, Swansea, United Kingdom}

\affiliation[b]{Swansea Academy of Advanced Computing, Swansea University (Bay Campus),\\
Fabian Way, Swansea SA1 8EN, United Kingdom}

\affiliation[c]{Higgs Centre for Theoretical Physics, School of Physics and Astronomy, The University of Edinburgh, \\
Peter Guthrie Tait Road, Edinburgh, EH9 3FD, United
Kingdom}

\emailAdd{2414441@swansea.ac.uk}
\emailAdd{e.j.bennett@swansea.ac.uk}
\emailAdd{gaurav.sinharay@swansea.ac.uk}

\abstract{
We present \texttt{hmcdj}, a new tool developed to ease the process of generating large numbers of gauge ensembles for a given theory. Building on top of the Grid library, hmcdj allows quick specification of lattice gauge theories in C++ with minimal boilerplate, and precise specification of per-ensemble parameters to be controlled at run time from a YAML input file. Additionally, it supports many quality-of-life features for managing large numbers of running jobs, including automatic tuning of the molecular dynamics step size to control the Metropolis acceptance rate; output of ILDG-compliant data; and integration with HPC schedulers to dynamically adapt the target trajectory count of a job to match its time limits. hmcdj is openly available under the GPLv2 license, and welcomes new contributions.}

\FullConference{
The 43rd International Symposium on Lattice Field Theory (Lattice 2026)

July 26 to August 1, 2026

University of Maryland, College Park, USA
}

\begin{document}
\maketitle

\section{Introduction}

Large scans over parameter spaces have appeared as a common theme in the research programme of the TELOS Collaboration.
Such scans are common in tuning the non-physical parameters to new actions, in studying theories at finite temperature, and in studies of theories beyond the Standard Model where a wide range of physical parameters must be investigated to characterise a theory.
Parameter scans in lattice field theory are intensive both in compute and human resources, requiring care to set up and effort to monitor and tune.

The large amount of human input on potentially hundreds of simulations running in parallel makes this process not only very expensive in human time but also very prone to human error.
The need to regenerate a set of ensembles,
perhaps because a forgotten parameter was held constant at the wrong value,
or a parameter to be varied was accidentally held fixed,
may cost a significant amount of computational resources,
and delay completion of a piece of work by months.

In order to prevent the above issues,
we suggest an increase in automation of the process.
Using deterministic processes with minimal inputs
reduces both the likelihood of human error
and the requirement for human time.
To minimise the number of exposed inputs in a given study,
we propose  a separation of concerns:
generally,
parameters requiring no change for a given study
may be compiled into a single executable,
which is used, unchanged, for all ensembles.
Parameters adjusted for every ensemble can be specified in an input file,
ensuring that they remain fixed for the entire generation process of that ensemble.
Some elements may need to be dynamic at run time;
we aim to automate the computation of these as much as possible.

To achieve the above we introduce \texttt{hmcdj}~\cite{hmcdj}, a lightweight C++ wrapper for the Grid data parallel C++ mathematical library~\cite{Boyle:2015tjk} for lattice field theory.
The TELOS Collaboration has  recently benefited from
the flexibility offered by the Grid environment,
in particular in its use GPU-based HPC resources,
by porting its original implementation of the Sp groups~\cite{Bennett:2017kga,Bennett:2023wjw},
first produced for HiRep~\cite{DelDebbio:2008zf}, into the Grid library~\cite{Bennett:2023gbe}.

The design of hmcdj is based on \textit{decks}
(executables defining a specific theory and computational setup),
which are supplied with parameters at runtime via \textit{tracks} (YAML input files).
The presence of the latter contributes towards reproducibility of the entire ensemble generation workflow.

Further, the wrapper provides a number of quality-of-life features which aim towards reducing the amount of human input and effort required to monitor, run and tune the simulations.
hmcdj is designed explcitly with use on High Performance Computing (HPC) platforms in mind, but is also compatible with use on smaller scale local systems.

hmcdj is extensively documented~\cite{hmcdjDocs} and we refer the reader to the documentation for further and updated information. 

\section{Motivation}

Development of hmcdj was concurrent with work on two prototype problems
(see Refs.~\cite{Sp4_Finite_Temperaure_with_Nf2,VerneyProvatas:lattice2026} and~\cite{Simonetti:2026jnx,Bennett:2026cev} 
respectively for details)
from the TELOS Collaboration's research programme~\cite{Bennett:2017kga,Bennett:2023wjw},
which may serve as useful illustrative use cases.
We will not go into detail into the physics of these projects,
but rather focus on the computational problems at hand, motivating the development and use of hmcdj.

\subsection{Sp(4) at finite temperature with fermions}

The goal of this project is to perform an initial scan of the phase space of the Sp(4) lattice gauge theory with $N_f=2$ Wilson-Dirac fermions, transforming in the fundamental representation~\cite{Sp4_Finite_Temperaure_with_Nf2,VerneyProvatas:lattice2026}. By varying the inverse gauge coupling $\beta$ and fermion bare mass $am_0$, the theory can transition between a confined and a deconfined regime.

Apart from the two parameters discussed above, in order to check the order of the phase transition multiple spatial lattice volumes $N_s^3$ are required at each point in parameter space. Further, to approach the continuum limit this study will require repetition at multiple values of the number of points in the temporal lattice dimension $N_t$.

We observe that a four dimensional parameter space has to be scanned. Potentially hundreds, even thousands, of ensembles are required. Considerable human effort is required to manage, monitor and tune this large number of jobs.

\subsection{Tuning of MDWF parameters for Sp(4)}

Ensemble generation using the Möbius Domain Wall Fermion formulation~\cite{Simonetti:2026jnx,Bennett:2026cev} requires tuning of a number of parameters, commonly found in the literature as $\alpha$, $a_5$, $m_5$ and $L_s$.
The tuning is performed with respect to minimising the residual mass $am_{\rm res}$.

Here we observe that for an optimal implementation,
for every value of the inverse gauge coupling $\beta$ a parameter scan in the above four dimensional parameter space is required before scanning of the parameters of physical interest can start.

Again, we require a very large number of simulations, using the same executable and different, user controlled, parameters for each run.

\section{Design}

Many existing HMC implementations adopt one of two paradigms:
in the first,
all,
or the majority of,
parameters are defined at compile time,
requiring compilation of a separate executable for each ensemble.
This works well for smaller libraries
that can quickly be compiled at the start of a job and give a small binary.
Grid~\cite{Boyle:2015tjk} outputs executables that are hundreds or thousands of megabytes in size,
and compilation adds a non-negligible overhead.
The opposing approach is to allow full specification of the theory of interest a run-time input file;
this allows maximum flexibility and reduces the amount of compilation needed,
but exposes a lot of unnecessary complexity to the user,
increasing the risk that a parameter is varied when it shouldn't be,
or is forgotten to be set to a value appropriate to the theory.

hmcdj aims to sit in between these two extremes,
allowing developers to relatively rigidly specify a theory in a ``deck''
(a single source file giving a single compiled executable),
but with the flexibilty to allow specific parameters to vary,
controlled by a ``track''
(an ensemble-specific input file,
of which there may be many in a given study).

hmcdj is designed as a thin C++ wrapper library for Grid.
Developers design \textit{decks}, Grid executables leaning on the core hmcdj library,
defining parameters which are to be supplied by users at runtime through \textit{tracks}, which act as input files.
Decks are minimal Grid executables, with the hmcdj library absorbing a large part of the boilerplate code required by Grid.
By restricting user freedom, when targeting a specific problem class---parameter scans---we aim to  restrict the freedom to make mistakes.

\subsection{Tracks}

Input to a deck is given in the form of a YAML~\cite{yaml} file, which we refer to as a \emph{track}. We choose YAML since it is designed to be both read and written by humans, and benefits from a mature ecosystem of parsers and tooling, eliminating any need to reinvent the wheel.

Tracks are deck specfic, and contain two separate namespaces:
a \verb|Global| namespace,
containing parameters common to any HMC study,
and a namespace specific to the deck that should run the track.
The former includes parameters such as the lattice dimensions,
checkpointing details,
and parameters for the molecular dynamics evolution.
The latter contains exactly the parameters that the developer has defined in the corresponding deck.
These are read automatically by the hmcdj library,
and passed to the relevant actions by the deck.

\subsection{Decks}

\textit{Decks} are template Grid executables, written using hmcdj's \verb|DJ| wrapper class.
Developers should be able to easily create new custom decks for their chosen theory, explicitly defining the custom parameters to be determined in the tracks.
hmcdj's aim is for the deck definition to be as concise as possible;
while there is still a little boilerplate code,
it is reduced compared with the equivalent implementation directly in Grid.

A compiled deck may (and indeed is expected to) be reused for all ensembles in a study;
parameters to change are set via the track,
which can be changed for each ensemble;
no interaction with the source code is required.

As tracks are designed to be deck specific, if the track namespace does not match the expected deck name, the executable will exit with failure;
likewise,
the structure is checked at run time,
so loading a malformed track
or one targeting an older version of the code
will fail,
rather than silently giving incorrect results.
For example,
attempting to run a track for a theory with dynamical matter
on the deck for pure gauge theory will fail at launch,
rather than generating an ensemble at the given value of $\beta$ but ignoring the fermion mass.
This allows such problems to be detected and fixed as soon as possible,
rather than being detectable only after analysing the output.

\section{Quality of life features}

We discuss some of the quality of life features offered by hmcdj, mainly aimed at reducing human input when performing a large number of HMC runs in parallel.

\subsection{Dynamic start}

Upon running, hmcdj either creates or searches through an ensemble directory, defined by the track parameters. The ensemble directory is determined relative to a base location. The base directory defaults to the user home directory, but can be overriden via an environment variable, for example when working in a project shared directory with multiple users contributing to a project. If the ensemble directory exists, hmcdj will check for the existence of saved configurations
and, if any are found, dynamically set the simulation starting type to resume the run.

The ensemble directory neatly contains gauge field and random number generator configurations along with logs produced by the runs. Gauge field configurations saved by the hmcdj checkpointer are in the ILDG 1.2 binary format\footnote{In future, we aim to also output ILDG ensemble and configuration metadata.}.

Since the production of a given ensemble typically requires many job submissions to an HPC facility
(due to wall time length restrictions typically imposed),
it is common to use array jobs or self-resubmitting jobs.
The dynamic start feature means that all such jobs can take the same form,
a single call to the relevant hmcdj deck,
rather than needing to have separate scripts for ``first job'' and ``subsequent jobs'',
or to have additional boilerplate to identify the trajectory  from which to resume.

\subsection{Slurm timing interface}

hmcdj interfaces with Slurm~\cite{yoo2003slurm}, checking whether the remaining time is enough to complete a trajectory. If the time left is estimated to not be enough to complete the next trajectory, hmcdj will save the current trajectory configuration and exit early with success.

This allows for easier automated resubmission of jobs, where hmcdj only exits with a successful exit code when it is appropriate for a subsequent job to automatically be submitted. This reduces the possibility of accidental tight resubmission loops that waste compute time; additionally, compute time is saved by not starting trajectories that will not complete before the job is killed.

\subsection{Acceptance rate tuning}

\begin{figure}
    \centering
    \includegraphics[width=0.9\linewidth]{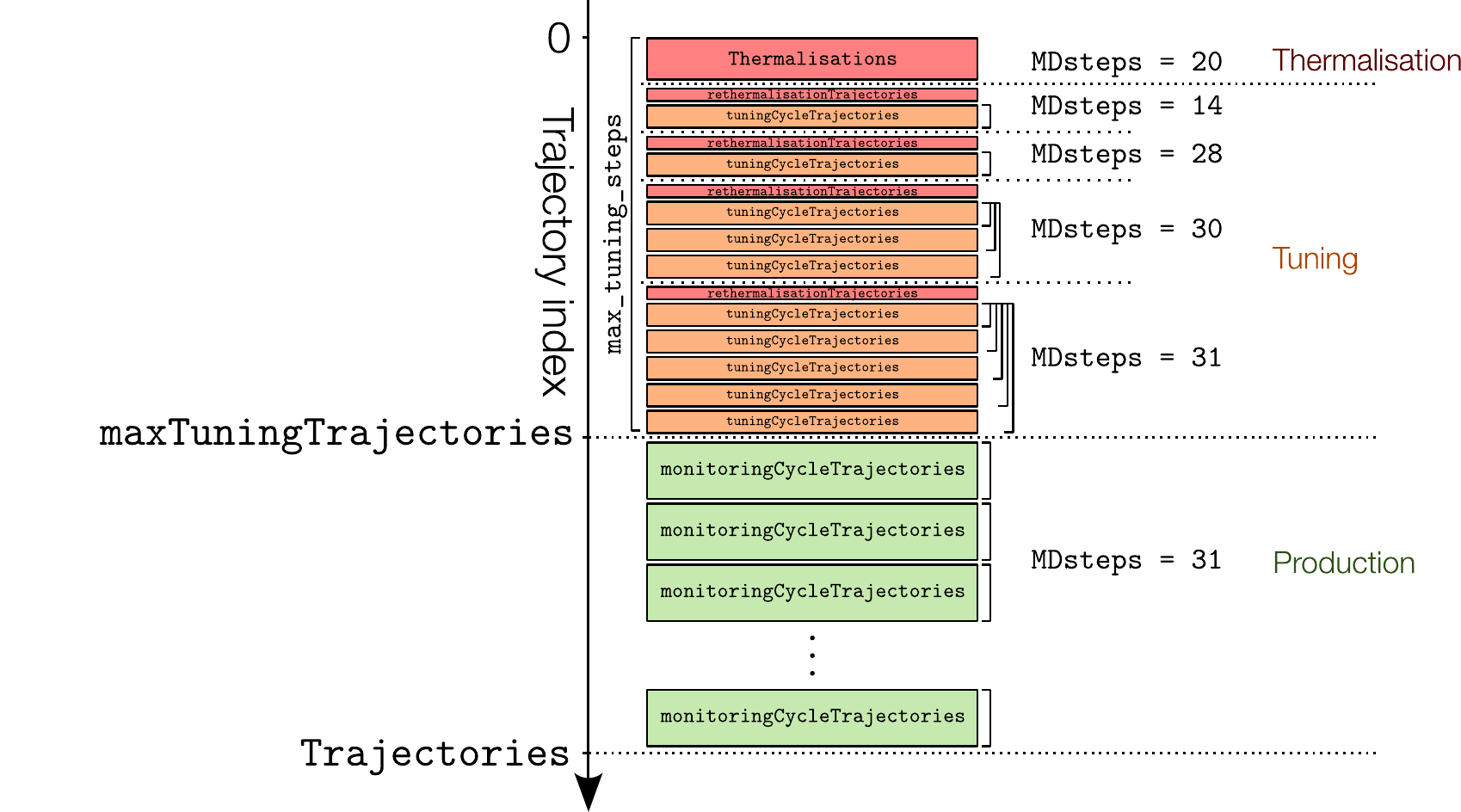}
    \caption{Typical acceptance rate tuning procedure with hmcdj.}
    \label{fig:acceptance_tuning}
\end{figure}

Tuning the molecular dynamics parameters---trajectory length and number of integration steps---to achieve a target acceptance rate, is an iterative process that requires monitoring and frequent human intervention. This can become extremely intensive in required human labour time when running hundreds of ensembles simultaneously. Incorrect monitoring and tuning can lead to wall and compute time wastage, with very long autocorrelation times or even the need to regenerate ensembles altogether.

In order to enable acceptance rate tuning at scale, hmcdj automates the process.
This is realised by monitoring the acceptance rate and modifying the molecular dynamics step size, in three separate phases. 
The parameters of the acceptance rate tuning are again controlled via the track in the \verb|Global| namespace.
We stress the design choice to keep the molecular dynamics trajectory length fixed to the user defined value, tuning only the number of integration steps.
A graphical representation of the acceptance rate tuning process can be found in Figure~\ref{fig:acceptance_tuning}.

In the initialisation phase, the Metropolis accept/reject step is not active and the acceptance rate is not monitored.

Next, the tuning phase is active for a number of chosen trajectories.
The acceptance rate is monitored in  intervals of specified trajectory length;
should it not fall within some user defined acceptance rate and tolerance range
(also taking into account the uncertainty on the estimate of the acceptance rate),
the number of molecular dynamics steps is modified.
Borrowing from Ref.~\cite{DelDebbio:2008zf}
the relationships of the expectation value of the difference in the Hamiltonian after the Molecular Dynamics evolution $\langle \Delta H \rangle$, first with the step size $\Delta\tau$, being $\langle \Delta H \rangle \propto (\Delta\tau)^4$, and second with the acceptance probability $p_{\rm acc} = \erfc(\sqrt{\langle \Delta H\rangle}/2)$, we estimate the target step size:

\begin{equation}
    \Delta\tau^{\mathrm{new}}
    = \Delta\tau^{\mathrm{old}}
    \sqrt{
    \frac{\erfc^{-1}(p_{\mathrm{acc}}^{\mathrm{target}})}{\erfc^{-1}(p_{\mathrm{acc}}^{\mathrm{old}})}
    }\ .
    \label{eq:mdsteps_update}
\end{equation}

In equation~\eqref{eq:mdsteps_update}, we clamp the acceptance rate within the range $[1/N_{\rm traj}, 1-1/N_{\rm traj}]$ with $N_{\rm traj}$ the current number of accumulated trajectories from which the acceptance rate is estimated.
This avoids cases where the step size would be set to zero or infinity should the current acceptance probability be estimated to be unity or zero respectively.

Since it is not acceptable to have the HMC step size vary through an ensemble,
once a specified number of trajectories have elapsed, hmcdj enters the monitoring phase.
An exit with error is called should the acceptance rate deviate outside the tolerance range.
This guarantees that for pathological cases,
an ensemble won't spend its entire generation failing to tune its acceptance,
instead alerting the user early that another fix must be employed.

\begin{figure}
    \centering
    \includegraphics[width=\textwidth]{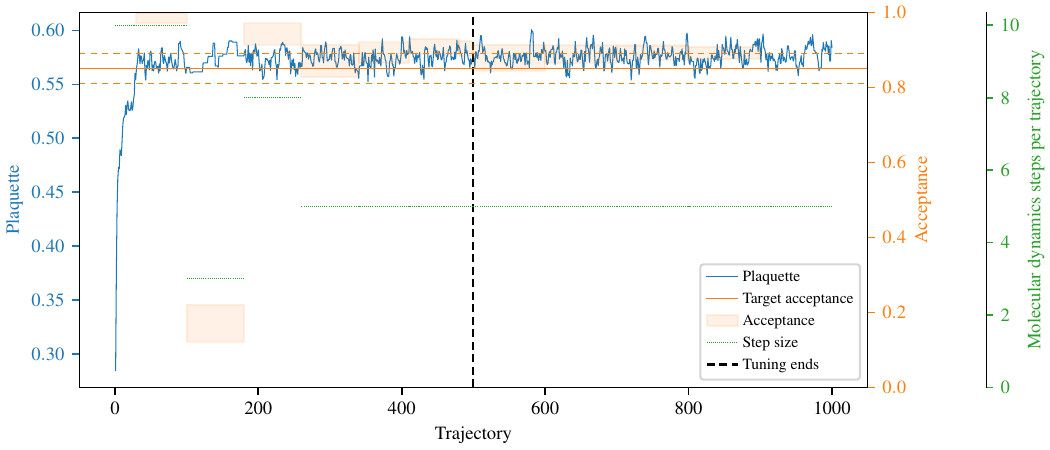}
    \caption{Example of acceptance rate tuning process.}
    \label{fig:tuning_example}
\end{figure}

Further technical details on the procedure are discussed in the documentation~\cite{hmcdjDocs}. An example case of the entire acceptance rate tuning procedure is visualised in Figure~\ref{fig:tuning_example}.

\section{Status and future work}

hmcdj is currently available under the GPLv2 license through the TELOS Collaboration's repository \cite{hmcdj}. Due to minor changes required from the upstream repository~\cite{UpstreamGrid}, hmcdj currently compiles when building against the TELOS Collaboration's fork of Grid~\cite{TELOSGrid}.
In this, we aim to make only non-invasive changes, with the goal that these be merged upstream once stable.
For compilation status against the upstream repository at time of reading, please check the hmcdj README~\cite{hmcdj}.

Compilation of hmcdj is implemented via autotools; requirements are a working installation of Grid and the yaml-cpp library~\cite{yaml-cpp}. The tool has been extensively documented~\cite{hmcdjDocs}, and includes a test suite, built on GoogleTest~\cite{gtest}, covering all major features.

Planned features for ongoing development include logging of on the fly observables into data files (removing the need to parse them from log files),
improvements to the output of metadata,
and improvements to running multiple Markov chains in parallel for a given parameter set.
More can be found on the Issues page of the repository~\cite{hmcdj}.

Development of further decks is problem specific and we hope that community members who are running similar workflows will find hmcdj useful and 
develop further decks for their own use cases.
Contributions,
either of such decks,
or of solutions to open issues,
are welcomed,
in accordance with our contribution guidelines.

\section{Acknowledgments}
We would like to thank Ryan Hill for the Grid executable template, upon which the autotools compilation of hmcdj was built, and advice on the implementation of required modifications to Grid. We would also like to thank Gianmarco Simonetti for useful discussions, user feedback and contributions to the development of the Möbius Domain Wall Fermion deck for Symplectic gauge theories with $N_f=2$ Dirac fermions in the fundamental representation. We also thank Maurizio Piai and Fabian Zierler for careful reading of and comments on this manuscript.

The work of E.~B. and G.~R. is supported by the STFC Research Software Engineering Fellowship EP/V052489/1 and has been supported in part by United Kingdom Research and Innovation (UKRI) via the Computational Science Centre for Research Communities (CoSeC), under Grant No.~UKRI497. The work of A.~V.-P. and E.~B. is supported by the STFC Consolidated Grant No.~ST/X000648/1. The work of A.~V.-P. is supported by a studentship funded by the Faculty of Science and Engineering, Swansea University, and by the School of Physics and Astronomy, The University of Edinburgh.

\subsection*{Open Access}

For the purpose of open access, the authors have applied a Creative Commons attribution
(CC BY) licence to any Author Accepted Manuscript version arising.

\subsection*{Data and Software Availability}

The data shown in Fig.~\ref{fig:tuning_example},
and the workflow used to create it,
are available at Ref.~\cite{data_release}.
hmcdj~\cite{hmcdj} and the version of Grid on which it depends~\cite{TELOSGrid}
are available via GitHub;
the specific versions used
(hmcdj commit \verb|7c9986d| and Grid commit \verb|f2a242b4|)
are included in Ref.~\cite{data_release}.
See also Ref.~\cite{Bennett:2025neg},
for a description of our approach to reproducibility and open science.

\bibliographystyle{unsrt}
\bibliography{ref}

\end{document}